\documentclass[prl,twocolumn,floatfix]{revtex4}

\usepackage{graphicx,amssymb,amsmath}
\usepackage{color,soul,bbold}

\begin{document}

\title{Confined Plasmons in Graphene Microstructures: Experiments and Theory}

\author{Jared H. Strait}
\email{jhs295@cornell.edu}
\author{Parinita Nene}
\author{Wei-Min Chan}
\author{Christina Manolatou}
\affiliation{School of Electrical and Computer Engineering, Cornell University, Ithaca, NY, 14853}
\author{Joshua W. Kevek}
\affiliation{Laboratory of Atomic and Solid State Physics and the Kavli Institute at Cornell for Nanoscale Science, Cornell University, Ithaca, NY 14853}
\author{Sandip Tiwari}
\affiliation{School of Electrical and Computer Engineering, Cornell University, Ithaca, NY, 14853}
\author{Paul L. McEuen}
\affiliation{Laboratory of Atomic and Solid State Physics and the Kavli Institute at Cornell for Nanoscale Science, Cornell University, Ithaca, NY 14853}
\author{Farhan Rana}
\affiliation{School of Electrical and Computer Engineering, Cornell University, Ithaca, NY, 14853}

\begin{abstract}
Graphene, a two-dimensional material with a high mobility and a tunable conductivity, is uniquely suited for plasmonics.  The frequency dispersion of plasmons in bulk graphene has been studied both theoretically and experimentally, whereas no theoretical models have been reported and tested against experiments for confined plasmon modes in graphene microstructures.  In this paper, we present measurements as well as analytical and computational models for such confined modes. We show that plsmon modes can be described by an eigenvalue equation. We compare the experiments with the theory for plasmon modes in arrays of graphene strips and demonstrate a good agreement.  This comparison reveals the important role played by interaction among the plasmon modes of neighboring graphene structures.
\end{abstract}

\maketitle

Graphene electronics and optoelectronics have emerged as fields of tremendous interest not only as improvements to existing technology, but also as platforms for completely novel devices.  A particularly interesting application of graphene is for plasmonic devices\cite{Ju11,HYan12,Rana08}, which manipulate charge density waves in the two-dimensional atomic sheet.  Graphene plasmons can have frequencies in the 1-100 terahertz range but wavelengths in the micron and sub-micron range\cite{Hwang07,Rana08,Seyller08}, enabling extreme confinement of electromagnetic energy.  In addition, plasmon frequencies in graphene can be tuned through electrostatic\cite{Ju11} or chemical\cite{HYan12} doping, making graphene plasmonics a unique platform for tunable terahertz sources, detectors, switches, filters, interconnects, and sensors. 

The dispersion of plamsons in bulk graphene has been obtained analytically\cite{Hwang07}, and the experimental results have been shown to agree well with the theoretical predictions\cite{Seyller08}. Plasmons can be confined in patterned graphene micro- and nano-structures such that only a discrete set of modes can oscillate\cite{Ju11,HYan12}. Such confined plasmon modes are of interest for device applications since, unlike bulk plasmons, they can couple directly to normally incident electromagnetic radiation. No analytical techniques have been reported that describe plasmon modes in arbitrary graphene structures and model interactions among plasmon modes of neighboring structures. In this paper, we present experimental and theoretical results for confined plasmon modes. We show that long wavelength plasmon modes in graphene microstructures can be described by an eigenvalue equation. The results obtained from the eigenvalue equation match well the results obtained using a finite-difference time-domain (FDTD) technique. By comparing the measured transmission spectra of interacting plasmon modes in an array of graphene strips with the theoretical results, we show that the theoretical models fit the experimental data very well.    

\begin{figure}[tbp]
	\centering
		\includegraphics[width=.35\textwidth]{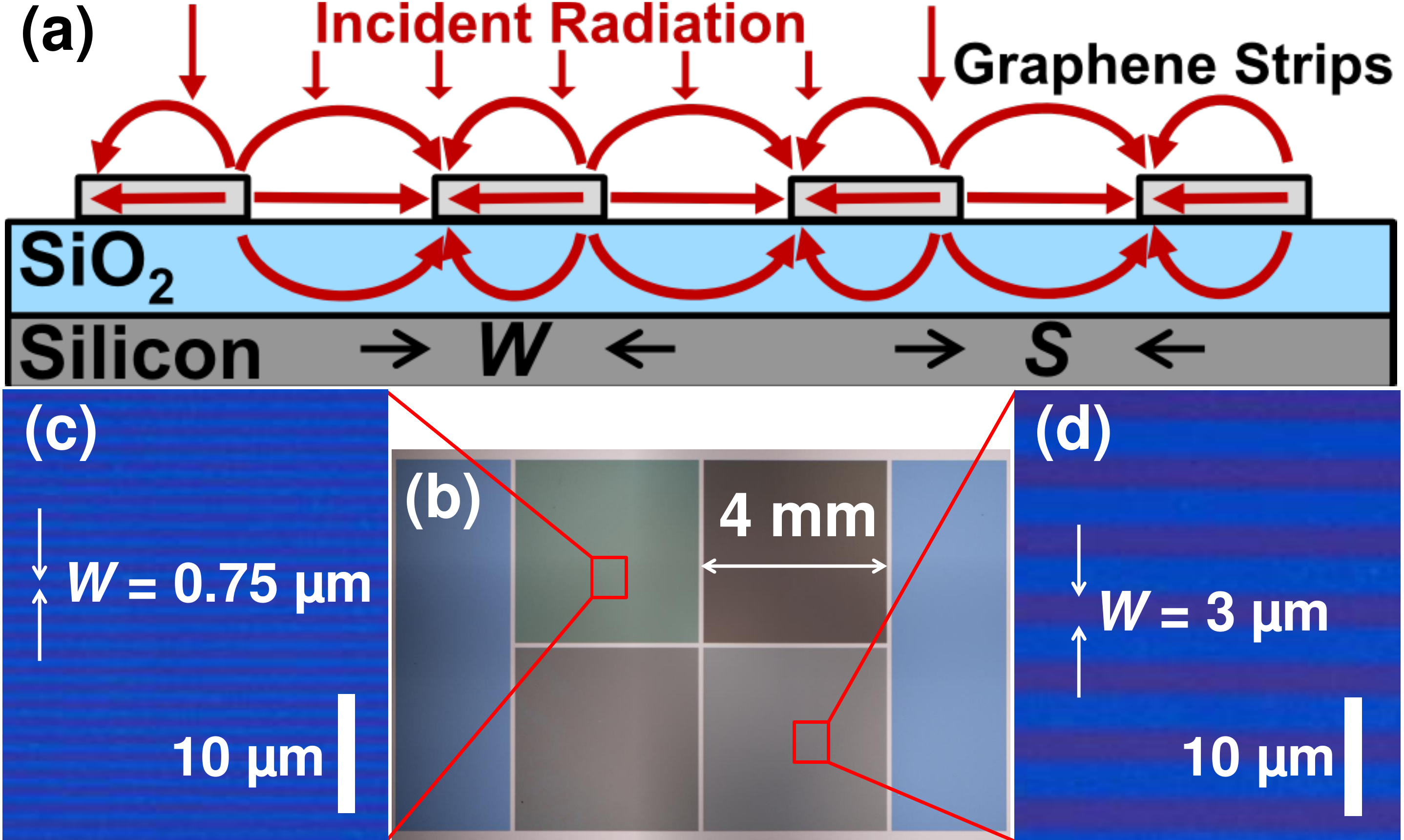}
	\caption{(a) A cross-section (not to scale) of an array of graphene strips with the electric field lines for the lowest plasmon supermode. (b) Optical micrograph of the sample with four regions of etched graphene strips (square regions in the center) and two reference regions.  (c,d) Bright-field optical micrographs (100X) detailing etched graphene strip arrays after the resist was removed.}
	\label{fig:Setup}
\end{figure}

%\section{Experiments: Far-IR Spectroscopy of Plasmon Modes in Arrays of Graphene Strips}
A cross-section of the structures considered in this work is shown in Fig.\ref{fig:Setup}(a), which shows an array of patterned graphene strips. The graphene used in our experiments was grown by chemical vapor deposition (Kevek Innovations 1'' System) on copper foils and transferred, as described by Li et al.\cite{Li09}, to high-resistivity silicon wafers ($>$10 k$\Omega$-cm) with $\sim$300 nm of thermally grown SiO$_2$.  Arrays of graphene strips of widths $W$ = 0.75, 1, 2, and 3 $\mu$m were patterned using standard photolithography followed by etching in an oxygen plasma (see Fig.\ref{fig:Setup}(b)).  For all devices, the strip width $W$ and the spacing $S$ between the strips were chosen to be equal. Graphene strips were doped using HNO$_3$\cite{Kasry10}. The doping density was estimated to be in the 4.5-5.0$\times10^{12}$ cm$^{-2}$ range using the Raman technique described by Das et al.\cite{Das08}.

\begin{figure}[tbp]
	\centering
		\includegraphics[width=.47\textwidth]{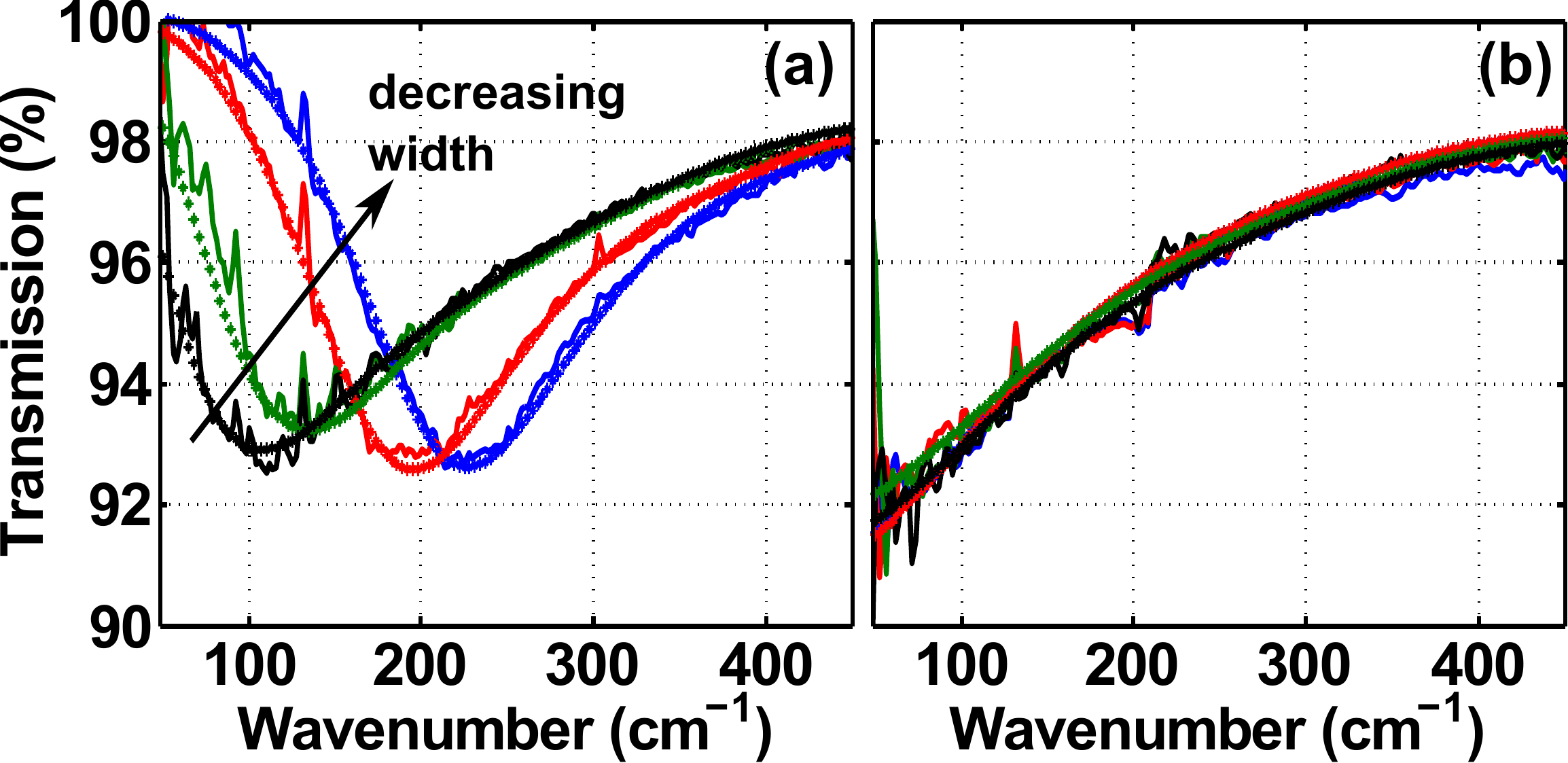}
	\caption{(Solid) Measured transmission of radiation polarized (a) perpendicular and (b) parallel to graphene strips is plotted for four different strip widths $W$ = 0.75, 1, 2, and 3 $\mu$m.  For all devices, the spacing $S$ between strips is equal to the width.  A bare SiO$_2$/Si substrate is used as reference.  (Dots) FDTD simulation fits to the measured results.  Extracted resonance frequencies are 226, 197, 135, and 112 cm$^{-1}$.}
	\label{fig:PerpParaFits}
\end{figure}

Plasmon resonances of arrays of graphene strips were measured at room temperature using Fourier-transform infrared spectroscopy (FTIR).  Fig.\ref{fig:PerpParaFits} shows the transmission spectra, of all four strip sizes, for polarizations perpendicular (a) and parallel (b) to the strips.  The transmission of incident radiation polarized parallel to the strips decreases monotonically at long wavelengths, showing the expected Drude-like frequency dependence.  There is no dependence on the strip width.  Transmission spectra of incident radiation polarized perpendicular to the strips show plasmon resonances\cite{Ju11,HYan12}.
 
%\section{Models and Simulations}
The measured transmission spectra can be described qualitatively using a damped harmonic oscillator model\cite{Nienhuys05,Strait09,Ju11,HYan12,Thongrattanasiri12},
\begin{equation}
\frac{T_{\xi}(\omega)}{T_{\textrm{ref}}(\omega)} = \left| 1 + \frac{\eta_{o} f \sigma(\omega\!=\!0)}{1 + n_{\textrm{sub}}}\frac{i\omega/\tau}{\omega^{2}-\xi\omega_{p}^{2} + i\omega/\tau} \right|^{-2} \label{eq:Tr}
\end{equation}
Here, $\eta_{o}$ is the free-space impedance, $\sigma(\omega\!=\!0)$ is the DC conductivity of bulk graphene, $f$ is the fill-factor of the strips, $n_{\textrm{sub}}$ is the refractive index of the substrate, $\tau$ is the Drude scattering time, $\omega_{p}$ is the plasmon frequency, and $\xi=0$ (or $1$) for incident radiation polarized parallel (or perpendicular) to the strips. The bulk plasmon frequency for small wavevectors is given by the expression\cite{Hwang07},
\begin{equation}
\omega_{p}(q) = \sqrt{\frac{\sigma(\omega\!=\!0)q}{2\epsilon_{\textrm{avg}} \tau}} \label{eq:Ex1}
\end{equation}
where $q$ is the magnitude of the plasmon wavevector and $\epsilon_{\textrm{avg}}$ is the average dielectric constant surrounding the graphene sheet.  Using $q = \pi/W$\cite{Xia12}, we find that Eq.\ref{eq:Ex1} significantly overestimates the plasmon frequencies compared to the experimental values. For example, using Eq.\ref{eq:Tr} to fit the transmission spectra of $W=1$ $\mu$m arrays for parallel polarizations ($\xi=0$), we find the average value of $\sigma(\omega=0)$ and $\tau$ to be 0.95 mS and 31.5 fs, respectively.  Eq.\ref{eq:Ex1} then gives a plasmon frequency of $\sim$245 cm$^{-1}$, which is significantly higher than the measured plasmon frequency of $\sim$197 cm$^{-1}$. Such a large error suggests that better models are needed to understand confined plasmon modes in patterned graphene structures.           

%\subsection{An Analytic Model for Graphene Plasmons}
We first present an analytic technique that captures the essential physics of the problem and results in an eigenvalue equation for the plasmon modes.  Assuming a Drude-like conductivity for graphene \cite{Dawlaty08}, we start with the time-derivative of the equation for the current density $\vec{K}$,
\begin{equation}
\frac{\partial^{2} \vec{K}}{\partial t^{2}}  + \frac{1}{\tau}\frac{\partial \vec{K}}{\partial t} = \frac{\sigma(\omega\!=\!0)}{\tau}\frac{\partial}{\partial t} \left( \vec{E}_{\rm inc} + \vec{E}_{d}\right) \label{eq:K1}
\end{equation}
Here, $\vec{E}_{inc}$ is the incident field and $\vec{E}_{d}$ is the depolarization field that results from the plasmon charge density.  Only field components in the plane of the graphene sheet are included in Eq.\ref{eq:K1}. Using the charge continuity equation and the Poisson equation, and ignoring retardation effects, the depolarization field can be related to the current density by,
\begin{equation} 
\frac{\partial  \vec{E}_{d}(\vec{r},t)}{\partial t} = \frac{-1}{4\pi\epsilon_{\textrm{avg}}}\int \textrm{d}^{2}\vec{r}' \bar{\bar{f}}(\vec{r}-\vec{r}') \cdot \vec{K}(\vec{r}',t) \label{eq:K2}
\end{equation}
The tensor $\bar{\bar{f}}(\vec{r}-\vec{r}')$ equals $\left[\mathbb{1} - 3 \vec{s}\otimes\vec{s}/| \vec{s} |^{2} \right]/| \vec{s} |^{3}$, where $\vec{s}=\vec{r}-\vec{r}'$.  $\bar{\bar{f}}(\vec{r}-\vec{r}')$ is related to the Green's function that relates the field to the polarization density and can be computed for more complicated geometries then considered in this work.  
We see from Eq.\ref{eq:K1} that if $(\sigma(\omega\!=\!0)/\tau)\partial \vec{E}_{d}/\partial t$ equals $-\omega_{p}^{2} \vec{K}$, then in the absence of $E_{\rm inc}$ and dissipation the current density will oscillate at the frequency $\omega_p$.
%Comparing Eq.\ref{eq:K1} and Eq.\ref{eq:K2}, we see that if  $(\sigma(\omega\!=\!0)/\tau)\partial \vec{E}_{d}/\partial t$ equals $-\omega_{p}^{2} \vec{K}$, then the effective complex conductivity, defined by $\vec{K}(\omega) = \sigma_{\rm eff}(\omega)\vec{E}_{inc}(\omega)$, is given by the expression\cite{Strait09}, $\sigma_{\rm eff}(\omega)/\sigma(\omega=0) = (i\omega/\tau)/(\omega^{2}-\omega_{p}^{2} + i\omega/\tau)$. The conductivity $\sigma_{\rm eff}$ then has the same frequency dependence as the effective conductivity given in Eq.\ref{eq:Tr} for $\xi=1$. 
Comparing with Eq.\ref{eq:K2}, it follows that the current density associated with the plasmon mode satisfies the following eigenvalue equation,
\begin{equation}
\frac{\sigma(\vec{r},\omega\!=\!0)}{4\pi\epsilon_{\textrm{avg}}\tau}  \int \textrm{d}^{2}\vec{r}' \bar{\bar{f}}(\vec{r}-\vec{r}') \cdot \vec{K}(\vec{r}')  = \omega_{p}^{2} \vec{K}(\vec{r}) \label{eq:K4}
\end{equation}
The above equation can be solved for the current densities $\vec{K_{m}}$ and frequencies $\omega_{pm}$ associated with the plasmon modes in any graphene structure. The modes satisfy the orthogonality condition, $\int d^{2}\vec{r} \vec{K}_{m}(\vec{r})\cdot \vec{K}_{p}(\vec{r})/\sigma(\vec{r},\omega\!=\!0) \propto \delta_{mp}$.  For the case of bulk plasmons, Eq.\ref{eq:K4} reproduces the result in Eq.\ref{eq:Ex1}. Solving the eigenvalue equation numerically for the case of a single infinitely long graphene strip, we obtain the following result for the frequency of the lowest two plasmon modes,
\begin{equation}
\omega_{p0} \approx \sqrt{\frac{\sigma(\omega\!=\!0) 1.156 }{\epsilon_{\textrm{avg}} \tau W}}, \,\,\,\, \omega_{p1} \approx \sqrt{\frac{\sigma(\omega\!=\!0) 2.751 }{\epsilon_{\textrm{avg}} \tau W}} \label{eq:Ex2}
\end{equation}

\begin{figure}[tbp]
	\centering
		\includegraphics[width=.48\textwidth]{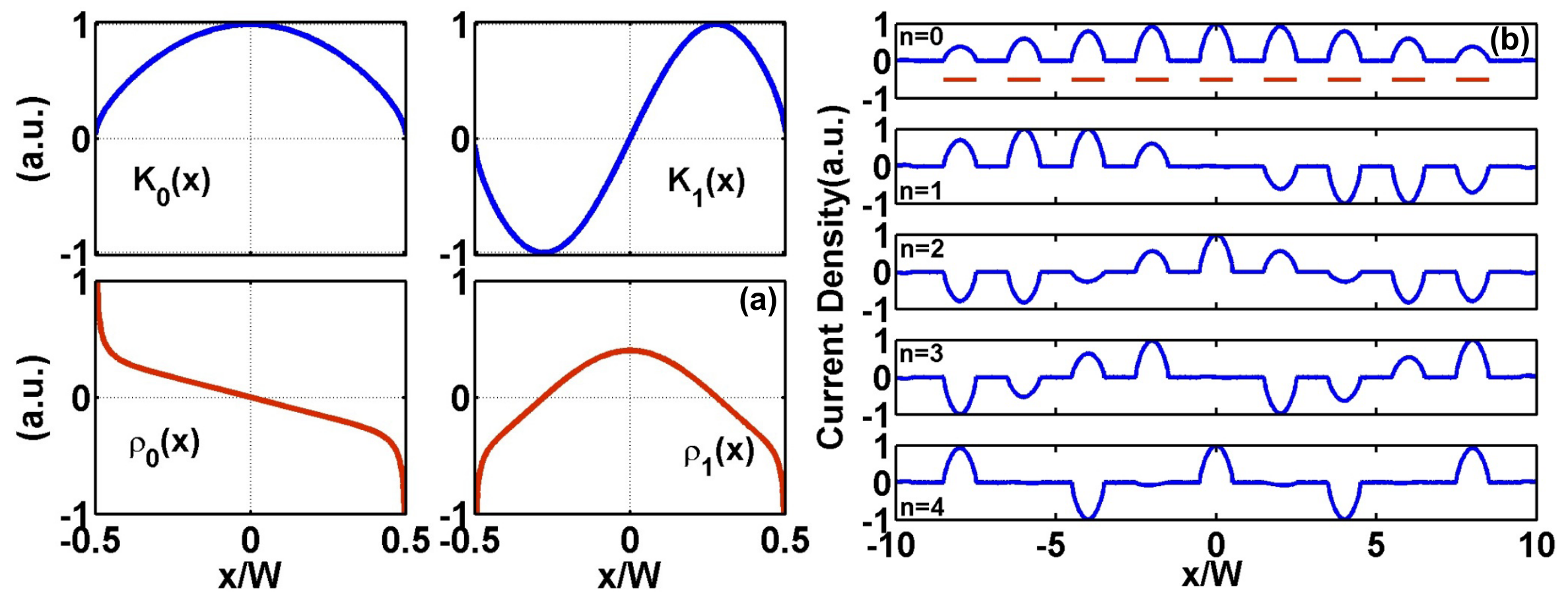}
	\caption{(a) The computed current densities $K(x)$ (top) and charge densities $\rho(x)$ (bottom) for the lowest two plasmon modes of a graphene strip are plotted. (b) The computed current densities are plotted for the first 5 supermodes of an array consisting of 9 graphene strips ($W$=$S$).  Locations of the strips are indicated by the red horizontal lines.} 
	\label{fig:supermodes}
\end{figure}

The computed current and charge densities for the lowest two plasmon modes are shown in Fig.\ref{fig:supermodes}(a). Although all even plasmon modes ($0,2,4...$) will couple with normally incident radiation, only the lowest plasmon mode will couple appreciably.  The scaling of the plasmon frequency with $1/\sqrt{W}$ is in perfect agreement with our data. For the case $W=1$ $\mu$m, using Eq.\ref{eq:Ex2} and the extracted values of $\sigma(\omega\!=\!0)$ and $\tau$, $\omega_{p0}$ is $\sim$211 cm$^{-1}$.  The eigenvalue equation more accurately models the plasma resonance than Eq.\ref{eq:Ex1} (with $q$=$\pi/W$), but it still overestimates the measured resonance frequency by $\sim$6.5\%.  We next address possible origins of this discrepancy. 

\begin{figure}[tbp]
	\centering
		\includegraphics[width=.48\textwidth]{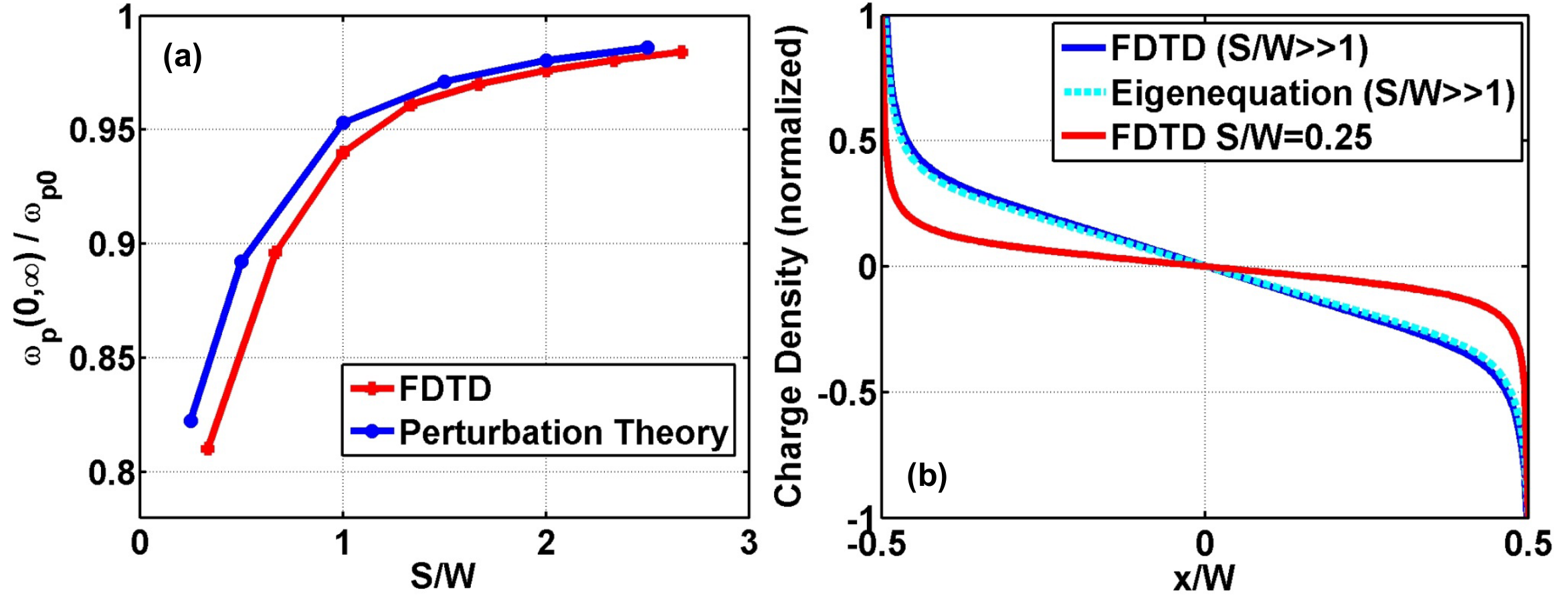}
	\caption{(a)  The frequencies of the lowest plasmon supermode $\omega_{p}(0,\infty)$ of an infinite graphene strip array calculated using perturbation theory and FDTD are plotted as a function of the ratio $S/W$.  (b) The computed charge densities for the lowest plasmon mode in an isolated graphene strip ($S/W\gg1$) and in an array of strips with $S/W=0.25$ are plotted. The charge densities were computed using FDTD ($S/W\gg 1$ and $S/W=0.25$) and the eigenvalue equation (Eq.\ref{eq:K4}) ($S/W\gg 1$).}
	\label{fig:fdtdvseig}
\end{figure}

Interactions among plasmon modes in neighboring strips can be included by solving Eq.\ref{eq:K4} using $\sigma(\vec{r},\omega\!=\!0)$ appropriate for an array of graphene strips.  Interactions lift the degeneracy among strips and result in a band of plasmon modes that are the supermodes of the array.  The computed current density for the lowest five supermodes of an array containing nine strips is shown in Fig.\ref{fig:supermodes}(b). Only the lowest supermode couples appreciably to the normally incident radiation, and it is the frequency of this supermode that is measured in our transmission experiments.  Unfortunately, the matrix eigenvalue equation obtained from Eq.\ref{eq:K4} is not sparse, so obtaining solutions for large arrays is computationally prohibitive.  Starting from the lowest plasmon mode of a single strip, perturbation technique can be used to obtain an expression for the frequency $\omega_{p}(n,N)$ and the current density $\vec{K}(n,N)(\vec{r})$ of the $n$th supermode ($n$=$0...N-1$) of an $N$-strip array,
\begin{equation}
\omega_{p}^{2}(n,N) \approx \omega_{p0}^{2} \left(1 - 2 \Delta_{1}\cos\left( \pi \frac{n+1}{N+1} \right) \right) 
\end{equation}
\begin{equation}
\vec{K}(n,N)(\vec{r}) \approx \sum_{j=1}^{N} \vec{K}_{0}(\vec{r}-j(S+W)\hat{x})\sin\left(\pi j \frac{n+1}{N+1} \right)
\end{equation} 
where $\Delta_{1}$ is the first nearest neighbor interaction parameter. Including second nearest neighbor interactions, $\omega_{p}^{2}(0,\infty)=\omega_{p0}^{2}\left(1 - 2 \Delta_{1}-2\Delta_{2} \right)$. $\Delta_{1}$ and $\Delta_{2}$ are given by the expression, 
\begin{equation}
\Delta_{\theta} = - \frac{\int \textrm{d}^{2}\vec{r} \int \textrm{d}^{2}\vec{r}' \vec{K}_{0}(\vec{r})\! \cdot\! \bar{\bar{f}}(\vec{r}-\vec{r}') \!\cdot\! \vec{K}_{0}(\vec{r}'\!-\!\theta (S\!+\!W) \hat{x})}{\int \textrm{d}^{2}\vec{r} \int \textrm{d}^{2}\vec{r}' \vec{K}_{0}(\vec{r}) \cdot \bar{\bar{f}}(\vec{r}-\vec{r}') \cdot \vec{K}_{0}(\vec{r}')}
\end{equation}
Fig.\ref{fig:fdtdvseig}(a) shows the the calculated values of $\omega_{p}^{2}(0,\infty)$ as function of the strip spacing $S$ assuming first and second nearest neighbor interactions. The plasmon frequency is reduced as a result of the interactions among neighboring strips. For $S$=$W$=1 $\mu$m, the values of $\Delta_{1}$ and $\Delta_{2}$ are $.035$ and $.009$, respectively, resulting in a $\sim$4.5\% decrease in the value of $\omega_{p}(0,\infty)$ compared to $\omega_{p0}$.  In this case, $\omega_{p}(0,\infty)\sim202$ cm$^{-1}$, which is closer to the measured $\sim$197 cm$^{-1}$ than $\omega_{p0}$ alone.  These results suggest that interactions cannot be ignored between nearby graphene plasmonic structures.

The eigenvalue equation does not include retardation effects, which could be important in the case of large arrays, and it also does not account for the discontinuity in the field at the oxide/silicon interface (screening by the silicon substrate).  A technique is needed that incorporates these effects, can be used to determine the accuracy of Eq.\ref{eq:K4}, and can also compute the measured transmission spectra more accurately than Eq.\ref{eq:Tr}.  For example, Eq.\ref{eq:Tr} predicts $T_{\xi=0}(\omega=0)=T_{\xi=1}(\omega=\omega_{p})$,  but the measured values in Fig.\ref{fig:PerpParaFits} differ by $\sim$1.5\%.  The discrepancy arises because the transmission through the gaps in the strip array cannot be modeled simply with a fill-factor $f$, especially when the incident radiation is polarized perpendicular to the strips and $S \leq W$.

%\subsection{A Finite-Difference Time-Domain (FDTD) Model for Graphene Plasmons}
In the FDTD method, Maxwell's equations are stepped in time. In order to model plasmons, we include an auxiliary equation for the graphene current density (Eq.\ref{eq:K1} without the extra time-derivative) and step the equations using Yee's leap-frogging algorithm\cite{FDTDBook}.  This approach naturally handles interactions and electromagnetic retardation.  A challenge in the FDTD technique is the range of length scales important to the problem.  The radiation frequencies of interest have free-space wavelengths extending up to 300 $\mu$m, but the corresponding plasmon wavelengths are on the order of 1 $\mu$m. Furthermore, it is important for the modeled graphene thickness to be much less than the plasmon wavelength.  Therefore, the length scales of importance span three orders of magnitude, necessitating a highly non-uniform mesh, with grid steps varying from $0.01-0.5$ $\mu$m.  The computational domain spans more than 200x200 $\mu$m$^2$, and is surrounded by perfectly-matched layer boundaries\cite{FDTDBook}. We use the values $4\epsilon_0$ and $12\epsilon_0$ for the THz dielectric constants of SiO$_2$ and Si, respectively. Plasmonic structures are excited at zero angle of incidence with a broadband (0.5-15 THz) pulse of electromagnetic radiation. The transmission spectra for fields polarized parallel and perpendicular to the strips are obtained by Fourier-transforming the time-domain transmitted pulse.  Values of $\sigma(\omega\!=\!0)$ and $\tau$ used in the simulations were iteratively improved until the simulated transmission spectra for both polarizations optimally fit the measured spectra.

The FDTD simulation results, shown in Fig.\ref{fig:PerpParaFits}, accurately fit the measurements.
Extracted values of $\sigma(\omega\!=\!0)$ and $\tau$ lie in the range 0.91-0.95 mS and 29.5-31.5 fs, respectively.  Using the expression for the graphene conductivity in Ref.\cite{Dawlaty08}, these values correspond to doping densities of $5.0-5.2\times10^{12}$ cm$^{-2}$, consistent with the densities determined using the Raman technique.  Small variations in the parameters across the CVD graphene sample are consistent with those measured by terahertz spectroscopy in similar samples\cite{Tomaino11}.  The ability of the FDTD technique to quantitatively fit the depth and width of the plasmon resonances, while also predicting their center frequencies to an accuracy within one percent, underscores its usefulness as a tool for modeling graphene plasmonic structures.

\begin{figure}[tbp]
	\centering
		\includegraphics[width=.48\textwidth]{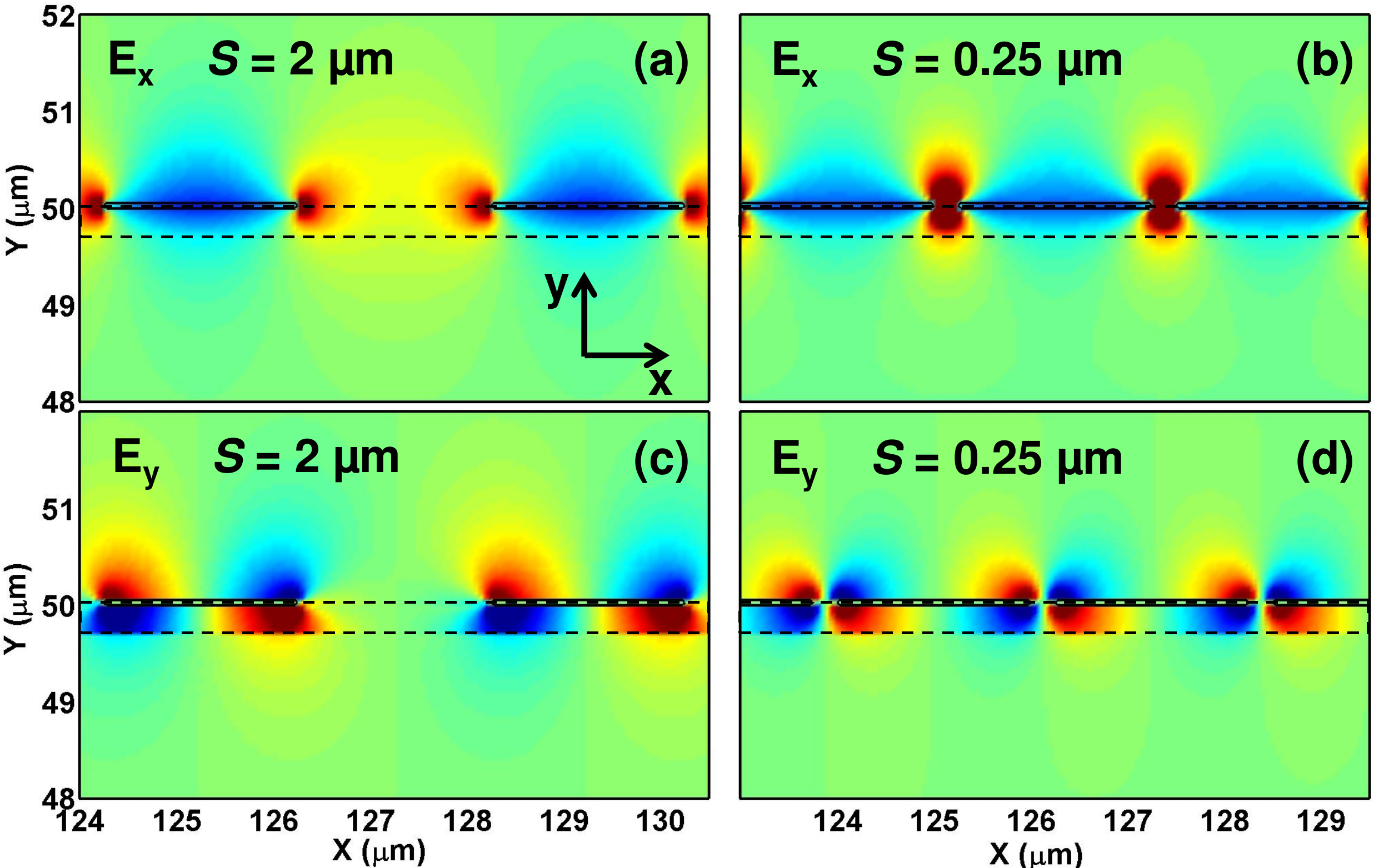}
	\caption{FDTD simulation results are shown for the $x$- (a,b) and $y$- (c,d) components of the electric fields for the lowest order plasmon supermode in two W = 2 $\mu$m arrays of graphene strips ($S$ = 2 $\mu$m (a,c), $S$=0.25 $\mu$m (b,d)). The dashed lines indicate the locations of the silicon/oxide and oxide/air interfaces.}
	\label{fig:EFields1}
\end{figure}

The computed $x$- and $y$-components of the electric field for the lowest plasmon supermode are shown in Fig.\ref{fig:EFields1} for arrays of graphene strips with two different spacings ($S$=2, 0.25 $\mu$m and $W$=2 $\mu$m).  The locations of the graphene strips are indicated by the thin black lines at $y=50$ $\mu$m.  The dashed lines indicate the locations of the silicon/oxide and oxide/air interfaces.  The field is highly localized near the graphene sheet, extending a distance on the order of the plasmon wavelength.  The discontinuity in the normal ($y$) component of the field at the silicon/oxide interface is also visible.  In contrast with the $S$ = 2.0 $\mu$m case, when $S$ = 0.25 $\mu$m, the field in the gaps between strips is stronger than the field in the center of the strips.  This effect helps to reveal the physical origins of the interaction between neighboring strips.  The plasmon charge density that accumulates at the edges of an isolated strip generates a depolarization field $\vec{E}_{d}$ with $\omega_{p0}^2\propto |\vec{E}_{d}|$.  In a strip array, this edge charge density is partially imaged on the neighboring strips, as depicted in Fig.\ref{fig:Setup}(a).  This effect increases the depolarization field in the gaps between strips but reduces the field within each strip.  Equivalently, the depolarization fields from neighboring strips are in-phase in the gaps between strips but out-of-phase in their centers. Therefore, $\omega_{p}(0,\infty)<\omega_{p0}$. In contrast, in the highest supermode of the array, the current density oscillations in neighboring strips are out of phase, so $\omega_{p}(N,\infty)>\omega_{p0}$.
 
FDTD simulation can serve to evaluate the perturbation theory, as shown in Fig.\ref{fig:fdtdvseig}. The two methods are found to agree when $S/W$$>$1, but not when $S/W$$\ll$1. For example, when $S$=$W$ in Fig.\ref{fig:fdtdvseig}(a), the FDTD calculated $\omega_{p}(0,\infty)$ is lower than $\omega_{p0}$ by $\sim$6.5\%, in contrast with the $\sim$4.5\% reduction predicted by the perturbation technique. This behavior can be understood by examining the plasmon charge density. Fig.\ref{fig:fdtdvseig}(b) shows the FDTD-computed charge density in a strip for the lowest-order supermode with $S/W$$\gg$1 and $S/W$=0.25.  Also shown is the charge density obtained by solving Eq.\ref{eq:K4} for $S/W$$\gg$1, which is nearly identical to the FDTD result.  But when $S/W$$\ll$1, the FDTD calculation reveals that the charge density is significantly modified as a result of interactions; the charge density increases near the strip edges to screen the fields of the neighboring strips. Since the perturbation theory assumed that the charge and current densities are unmodified from the lowest plamon mode of an isolated strip, the results became inaccurate when $S/W$$\ll$1.  The good agreement obtained between the FDTD method and the analytic model for $S/W$$>$1 suggests that retardation effects do not play a significant role in the structures considered in this work.

%\section{Conclusion}
To conclude, we presented experimental and theoretical results for the confined plasmon modes in graphene microstructures.  We presented an analytic model which captures the essential physics and gives an eigenvalue equation for computing plasmon modes.  We also presented a universally applicable FDTD technique. The theoretical models presented show good agreement with the measurements, and demonstrate the importance of interactions in plasmonic structures. Recently, numerical and analytical results using other approaches have been reported for graphene with periodically modulated conductivity~\cite{Peres12,Peres13, Ferreira12, Fallahi12} and graphene disk arrays\cite{Thongrattanasiri12} in a slightly different context. The present work, to the best of our knowledge, is the first time that theoretical and numerical models have been presented and tested against experiments for confined plasmon modes in graphene microstructures. The techniques presented in this paper can be used to understand, model, and design complex graphene plasmonic structures for applications ranging from IR detectors and chemical sensors to plasmonic radiation sources, oscillators, modulators, and metamaterials. 

The authors would like to acknowledge helpful discussions with Michael G. Spencer and support from CCMR under NSF grant number DMR-1120296, AFOSR-MURI under grant number FA9550-09-1-0705, ONR under grant number N00014-12-1-0072, and the Cornell Center for Nanoscale Systems funded by NSF.

%Peres12,Peres13
%Thongrattanasiri12,Ferreira12

\bibliography{FDTDGrPlasBib}

\end{document}